\documentclass[nofootinbib,superscriptaddress,aps,preprint,prc]{revtex4}
\usepackage[utf8]{inputenc}
\usepackage{amsmath}
\usepackage{mathrsfs}
\usepackage{amssymb}
\usepackage{graphicx}
\usepackage{graphics}
\usepackage[small,compact]{titlesec}
\usepackage{xspace}
\usepackage{colordvi}
\usepackage{color}
\usepackage{stackrel}
\usepackage{hyperref}
\usepackage{slashed}

\usepackage{ulem}

\renewcommand{\emph}[1]{\textit{#1}}

\newcommand{\ya}{\gamma_{\mathrm{AVE}}}
\newcommand{\tso}{{}^{3}\!S_{1}}
\newcommand{\osz}{{}^{1}\!S_{0}}

\newcommand{\vect}[1]{\vec{\mathbf{#1}}}

\newcommand{\comment}[1]{}

\newcommand{\jjvHe}{{}^3\mathrm{He}}
\newcommand{\jjvH}{{}^3\mathrm{H}}

\newcommand{\eftnopi}{EFT($\slashed{\pi}$)\xspace}

\newcommand{\G}{\mathcal{G}}
\newcommand{\Nt}{\hat{N}^{T}}
\newcommand{\Nd}{\hat{N}^{\dagger}}
\newcommand{\N}{\hat{N}}
\newcommand{\Gt}{\widetilde{\mathcal{G}}}

\newcommand{\NN}{$N\!N$\xspace}
\newcommand{\Nc}{$N_{C}$\xspace}

\makeatletter

\newcommand{\Rmnum}[1]{\expandafter\@slowromancap\romannumeral #1@}
\newcommand{\vast}{\bBigg@{4}}
\newcommand{\Vast}{\bBigg@{5}}
\makeatother

\begin{document}

\title{Three-nucleon bound states and the Wigner-SU(4) limit}

\author{Jared Vanasse}
\email{vanasse@ohio.edu}
\affiliation{Institute of Nuclear and Particle Physics and Department of Physics and Astronomy Ohio University, Athens OH 45701, USA
}

\author{Daniel R. Phillips}
\email{phillips@phy.ohiou.edu}
\affiliation{Institute of Nuclear and Particle Physics and Department of Physics and Astronomy Ohio University, Athens OH 45701, USA
}

\date{\today}

\begin{abstract}
We examine the extent to which the properties of three-nucleon bound states are well-reproduced in the limit that nuclear forces satisfy Wigner's SU(4) (spin-isospin) symmetry. To do this we compute the charge radii up to next-to-leading order (NLO) in an effective field theory (EFT) that is an expansion in powers of $R/a$, with $R$ the range of the nuclear force and $a$ the nucleon-nucleon (\NN) scattering lengths. In the Wigner-SU(4) limit, the triton and Helium-3 point charge radii are equal. At NLO in the range expansion both are $1.66$~fm. Adding the first-order corrections due to the breaking of Wigner symmetry in the \NN scattering lengths gives a $\jjvH$ point charge radius of $1.58$~fm, which is remarkably close to the experimental number, $1.5978\pm0.040$~fm~\cite{Angeli201369}.  For the $\jjvHe$ point charge radius we find $1.70$~fm, about 4\% away from the experimental value of $1.77527\pm0.0054$~fm~\cite{Angeli201369}.  We also examine the Faddeev components that enter the tri-nucleon wave function and find that an expansion of them in powers of the symmetry-breaking parameter converges rapidly. Wigner's SU(4) symmetry is thus a useful starting point for understanding tri-nucleon bound-state properties. 
\end{abstract}


\maketitle

\section{Introduction}

Quantum-mechanical systems in which the two-particle potential is short-ranged, and the two-body scattering length is large compared to that range, share ``universal" features~\cite{Braaten:2004rn}.  The most striking of these is the Efimov effect; the existence of an infinite series of three-body bound states. In the “unitary limit" the scattering length $|a|\rightarrow \infty$, and the three-body problem exhibits discrete scale invariance, with states in the Efimov tower related to one another through a rescaling of co-ordinates by a factor that is $e^{\pi/s_0}=22.7$~\cite{Efimov:1970zz,Efimov:1971zz} for the equal-mass case. The existence of two states related by this Efimov ratio has recently been demonstrated for Cesium atoms near a Feshbach resonance---i.e. essentially in the unitary limit~\cite{Huang:2014}---and for clusters of Helium atoms that have a large, but finite, two-body scattering length~\cite{Kunitski:2015qth}.

Both of these systems consist of bosons, whereas the particles that make up nuclei are spin one-half fermions of two different isospins. This means that---even in the approximation that $S$-wave interactions dominate the formation of the three-nucleon bound state---\NN (nucleon-nucleon) interactions in two different channels, the ${}^1$S$_0$ and the ${}^3$S$_1$, contribute to the binding of the three-nucleon system.  Nevertheless, the Efimov effect also occurs for three nucleons~\cite{Efimov:1981kfh}: the virtual state in doublet S-wave neutron-deuteron scattering becomes an excited Efimov state of the triton in appropriate limits~\cite{Adhikari:1982zzb,Frederico:1988zza,Kievsky:2015dtk}. Most recently, Kievsky and Gattobigio studied the physics of the three-nucleon bound state with model Gaussian potentials~\cite{Kievsky:2015dtk}, showing that Efimov states appear in the three-nucleon spectrum as the ${}^1$S$_0$ and ${}^3$S$_1$ scattering lengths tend towards the unitary limit.  They argued that this means the triton is inside the ``Efimov window" in that its structure is governed by `a few control parameters, [such] as the two-body energies and scattering lengths', i.e. it can be described within the context of few-body universality. 

An effective field theory (EFT) with only short-range interactions provides a systematic way to organize the treatment of three-body states in this universal/Efimov-window regime. It exploits the hierarchy of scales $R \ll |a|$, and in nuclear physics it is known as the pionless EFT (\eftnopi)~\cite{vanKolck:1998bw,Kaplan:1998tg,Kaplan:1998we,Birse:1998dk}. At leading order (LO) in \eftnopi the particles interact via zero-range forces, whose strengths are tuned to reproduce, e.g., the ${}^1$S$_0$ and ${}^3$S$_1$ scattering lengths. At higher orders corrections to two-body observables due to the finite effective ranges, $r$, can be computed in perturbation theory~\cite{Chen:1999tn}, with a nominal expansion parameter of $r/a \approx 30$\% in the ${}^3$S$_1$ channel. 

 The leading-order equations for the triton in this EFT were worked out in Ref.~\cite{Bedaque:1999ve}, and it was quickly apparent that those equations are equivalent to the (single) equation for bosons~\cite{Bedaque:1998kg,Bedaque:1998km} in the limit that the ${}^3$S$_1$ and ${}^1$S$_0$ scattering lengths are equal, i.e., if the \NN interaction displays a Wigner-SU(4) spin-isopsin symmetry~\cite{Bedaque:1999ve,Mehen:1999qs}. That equation, known as the STM (Skornyakov-Ter-Martirosian~\cite{Skornyakov}) equation, must be regulated.
 Using a momentum-space cutoff $\Lambda$ its solution is sensitive to the value of $\Lambda$, i.e. to short-distance physics in the three-body system;  the STM equation does not posses a unique solution in the limit $\Lambda\to\infty$~\cite{danilov:1961}. These problems can be removed by adding a three-body force to the EFT at leading order~\cite{Bedaque:1998km}. The three-body force prevents Thomas collapse~\cite{Thomas:1936}.
 
The leading-order EFT  calculation recovers the prediction of the Efimovian spectrum in the unitary limit and also permits straightforward extension of that result to finite scattering lengths---and to finite, and different, $S=0$ and $S=1$ scattering lengths in the nuclear-physics case.  This reproduces findings of Efimov~\cite{Efimov:1970zz,Efimov:1971zz,Efimov:1981kfh} and others~\cite{Adhikari:1982zzb,Frederico:1988zza} for zero-range forces. Crucially, the LO three-body force in the three-nucleon problem is Wigner-SU(4) symmetric~\cite{Mehen:1999qs}---even for the situation where the $S=0$ and $S=1$ channels exhibit a different scattering length; Wigner-SU(4)-anti-symmetric three-body forces do not enter the EFT until much higher orders in the expansion~\cite{Griesshammer:2005ga,Barford:2004fz,Birse:2005pm}. Higher orders in the $R/a$ expansion are calculated by considering perturbative corrections to three-body observables due to the finite range of the nuclear force. EFT calculations at next-to-leading (NLO) and next-to-next-to-leading-order (NNLO) in the range appeared in Refs.~\cite{Platter:2008cx,Ji:2010su,Ji:2011qg,Ji:2012nj} (for three bosons) and \cite{Hammer:2001gh,Bedaque:2002yg,Vanasse:2013sda} (for the three-nucleon system). Most recently, Vanasse has shown that the triton point charge radius is well described within \eftnopi, obtaining $\langle r_{\jjvH}^{2} \rangle_{\rm pt}=1.14 + 0.45 + 0.03=1.62$~fm at leading order and for NLO and NNLO corrections~\cite{Vanasse:2015fph}. The NLO and NNLO results agree with the experimental value of $1.5978\pm0.040$ fm~\cite{Angeli201369}. While the NLO correction is sizable, the excellent agreement and reasonable convergence pattern support the contention of Ref.~\cite{Kievsky:2015dtk} that the triton is within the purview of few-body universality. 

In this paper we use \eftnopi to answer the question of how relevant Wigner-SU(4) symmetry is to the physics of both the triton and ${}^3$He. Naively the \NN system seems far from the Wigner-SU(4) limit: the deuteron binding momentum is 45 MeV, while the corresponding scale in the ${}^1$S$_0$ channel, the inverse of the $\osz$ neutron-proton ($np$) scattering length, is $1/a_{np}^{S=0}=-8.3$ MeV. Thus the parameter that governs Wigner-SU(4) breaking:
\begin{equation}
\delta \equiv \frac{1}{2} (1/a_{np}^{S=1} - 1/a_{np}^{S=0})
\end{equation}
is not small compared to the average of $1/a_{np}^{S=1}$ and $1/a_{np}^{S=0}$. However, we shall see that an expansion around the Wigner-SU(4) limit, where $\delta=0$, converges well. The triton binding energy changes by only 0.8 MeV due to Wigner-SU(4) breaking, and the triton charge  radius in the Wigner-SU(4) limit is $1.66$~fm at NLO in \eftnopi, quite close to the average of the experimental $\jjvH$ and $\jjvHe$ point charge radii. Perhaps most tellingly, the Wigner-SU(4)-odd component of the triton wave function is less than 10\% the size of the SU(4)-even part, which implies that an expansion around the Wigner-SU(4) limit will be successful for all trinucleon bound-state observables---or at least for all observables that do not vanish in that limit.

Wigner-SU(4) (spin-isospin) symmetry has had considerable phenomenological success in nuclear physics, ever since, in 1937, Wigner classified nuclear states according to their SU(4) representation in order to explain the pattern of nuclear masses up to $A \approx 40$~\cite{Wigner:1936dx}. Subsequently he worked out the consequences of such a symmetry for nuclear beta decays~\cite{Wigner:1939zz}. The ``Wigner super-multiplet theory" was later applied to inelastic electron scattering from, and muon capture on, ${}^{12}$C and ${}^{16}$O~\cite{Foldy:1964,Walker:1966zz,deForest:1965}; the particle-hole states were usefully classified according to Wigner-SU(4), thereby explaining the existence of a family of giant resonances in these nuclei. 

We note that the presence of Wigner-SU(4) symmetry in the three-nucleon problem is a weaker condition than that the three-nucleon problem exhibit the unitary ($|a| \rightarrow \infty$) limit. The unitary limit may be relevant for few-nucleon systems in large magnetic fields~\cite{Detmold:2015daa} or in a version of QCD with slightly larger and unequal up- and down-quark masses~\cite{Braaten:2003eu}. Recently, K{\"o}nig {\it et al.} have argued that the binding energies of the $A=3$ and $A=4$ systems can be understood both qualitatively and quantitatively via an expansion around the unitary limit. We will comment specifically on this idea in Sec.~\ref{sec:unitary}. 
 In the Wigner-SU(4) limit the four \NN scattering lengths $a_{nn}$, $a_{pp}$, $a_{np}^{S=0}$, and $a_{np}^{S=1}$ are all equal, but could be finite. Efimovian towers can still occur for finite scattering lengths (e.g. the helium trimers), but they are related by a scaling factor which is smaller than the 22.7 that applies for equal masses when $|a| \rightarrow \infty$. 
In this situation the equations for the triton are those for a two-neutron halo with a neutron-core scattering length equal to the neutron-neutron scattering length~\cite{Canham:2008jd}. Therefore the Wigner-SU(4) limit not only connects the trinucleons to the three-boson systems being investigated experimentally in Innsbruck~\cite{Huang:2014}, Frankfurt~\cite{Kunitski:2015qth}, and elsewhere, it also permits us to understand the triton as the lightest two-neutron halo. 

Our discussion of this connection proceeds as follows.  In Sec. \ref{sec:two-body} we introduce the basic formalism for Wigner-SU(4) symmetry and its breaking in the two-body sector, while Sec. \ref{sec:three-body} introduces this formalism in the three-body sector.  Sections \ref{sec:binding}, \ref{sec:radii}, and \ref{sec:expansion} discuss the effects of Wigner-SU(4) symmetry and its breaking on binding energy, charge and matter radii, and triton vertex functions.  In Sec.~\ref{sec:unitary} we examine the values obtained for three-nucleon charge radii in the unitary limit and in Sec. \ref{sec:conclusions} we conclude.

\section{\label{sec:two-body}Wigner-SU(4) Symmetry in the Two-Body Sector}

The LO \NN interaction in \eftnopi can be written as~\cite{Weinberg:1991um}
\begin{equation}
\mathcal{L}_{2}=-\frac{1}{2}C_{0}^{T}\Nd\boldsymbol{\sigma}_{i}\N\Nd\boldsymbol{\sigma}_{i}\N-\frac{1}{2}C_{0}^{S}\Nd\N\Nd\N.
\label{eq:Weinbergbasis}
\end{equation}
A Wigner transformation $\N\to \hat{U}\N$ is a simultaneous transformation under spin and isospin given by the operator $\hat{U}=e^{i\alpha_{\mu\nu}\sigma_{\mu}\tau_{\nu}}$, where $\sigma_{\mu}=\{1,\boldsymbol{\sigma}_{i}\}$ and $\tau_{\nu}=\{1,\boldsymbol{\tau}_{a}\}$ are four vectors with $\mu,\nu=0,1,2,3$ and $i,a=1,2,3$.  The determinant of $\hat{U}$ is equal to one and $\alpha_{\mu\nu}$ is a $4\times 4$ matrix of real numbers~\cite{Wigner:1936dx,Mehen:1999qs}, with $\alpha_{00}=0$.  It is immediately obvious that the $C_{0}^{S}$ term is invariant under a Wigner transformation while the $C_{0}^{T}$ term is not.  Thus \eftnopi is Wigner symmetric at LO if and only if $C_{0}^{T}=0$.  The LO $NN$ interaction can also be written in the partial-wave basis yielding
\begin{equation}
\mathcal{L}_{2}^{PW}=-C^{(\tso)}_{0}\left(\Nt P_{i}\N\right)^{\dagger}\!\!\left(\Nt P_{i}\N\right)-C^{(\osz)}_{0}\left(\Nt \bar{P}_{a}\N\right)^{\dagger}\!\!\left(\Nt \bar{P}_{a}\N\right),
\end{equation}
where $P_{i}=\frac{1}{\sqrt{8}}\sigma_{2}\sigma_{i}\tau_{2}$ ($\bar{P}_{a}=\frac{1}{\sqrt{8}}\sigma_{2}\tau_{2}\tau_{a}$) projects out the spin-triplet iso-singlet (spin-singlet iso-triplet) combination of nucleons.  Parameters in Eq.~(\ref{eq:Weinbergbasis}) can be related to parameters in the partial wave basis via~\cite{Mehen:1999qs}
\begin{equation}
C^{(\osz)}_{0}=C_{0}^{S}-3C_{0}^{T}\quad,\quad C^{(\tso)}_{0}=C_{0}^{S}-C_{0}^{T},
\end{equation}
so the condition $C_{0}^{T}=0$ for Wigner-SU(4) symmetry is equivalent to $C^{(\osz)}_{0}=C^{(\tso)}_{0}$ in the partial-wave basis.  At LO in the \eftnopi power counting the \NN scattering amplitude is given by an infinite sum of bubble diagrams~\cite{Kaplan:1998tg,Kaplan:1998we}. Fitting to the $^3$S$_1$ ($^1$S$_0$) bound (virtual bound) state pole gives
\begin{equation}
C^{(\tso)}_{0}=\frac{4\pi}{M_{N}}\frac{1}{\gamma_{t}-\mu}\quad,\quad C^{(\osz)}_{0}=\frac{4\pi}{M_{N}}\frac{1}{\gamma_{s}-\mu},
\end{equation}
for the low-energy constants (LECs) in the partial-wave basis. (The scale $\mu$ comes from using the power divergence subtraction scheme with dimensional regularization~\cite{Kaplan:1998tg,Kaplan:1998we}.)  If $\mu\gg \gamma_{t},\gamma_{s}$ then Wigner-SU(4) symmetry is approximate in the \NN system.  However, if $\gamma_{t}=\gamma_{s}$  then Wigner-SU(4) symmetry is exact for the \NN system at LO.  $\gamma_{t}=45.7025$~MeV and $\gamma_{s}=-7.890$~MeV~\cite{Griesshammer:2004pe} correspond to the momenta at which poles of the \NN scattering amplitude occur in the ${}^3$S$_1$ and ${}^1$S$_0$ channels, respectively. At LO in the \eftnopi expansion they are equal to $1/a_{np}^{S=1}$ and $1/a_{np}^{S=0}$~\cite{Chen:1999tn,Beane:2000fx,Griesshammer:2004pe}.  Since $\gamma_s \neq \gamma_t$ Wigner-SU(4) symmetry is not exact.  We will explore the extent to which an expansion in powers of $\gamma_s - \gamma_t$ gives access to the properties of three-nucleon bound states.  

Up to NLO in the EFT expansion the Wigner-SU(4) symmetric limit is attained if all effective-range expansion parameters occurring up to that order are equal in the $\tso$ and $\osz$ channels. This results in equal Lagrangian parameters in the $\osz$ and $\tso$ channels, thus guaranteeing symmetry of the Lagrangian under Wigner-SU(4) transformations. Tensor interactions complicate the definition at higher orders. But at NLO this means that Wigner-SU(4) symmetry is satisfied if and only if the $\osz$ and $\tso$ channels have equal scattering lengths and effective ranges.

\section{\label{sec:three-body}Wigner-SU(4) Symmetry in the Three-Body Sector}

\begin{figure}[hbt]
\includegraphics[width=110mm]{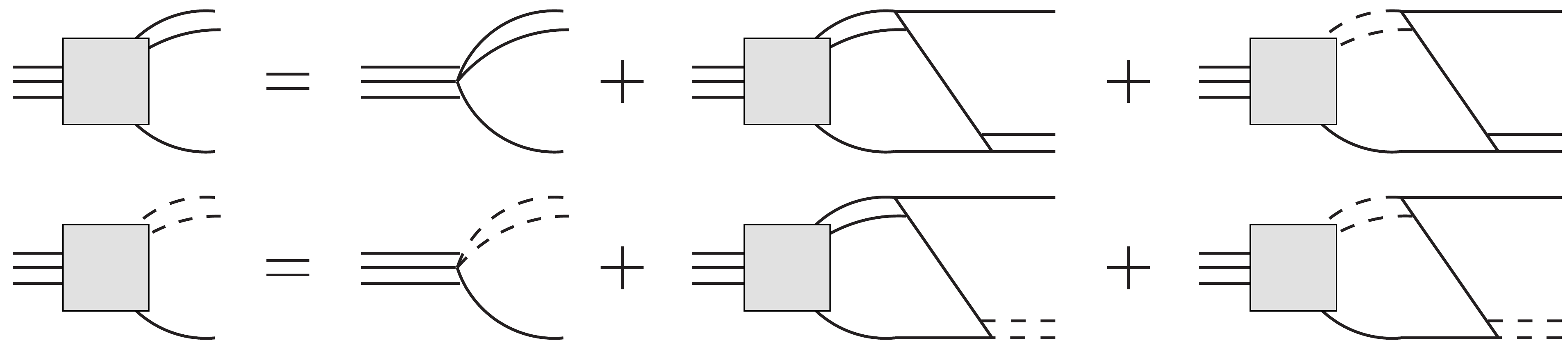}
\caption{\label{fig:GirrLO}Set of coupled integral equations for the LO tri-nucleon vertex function.  Single lines are nucleons, double lines $\tso$ dibaryons, double dashed lines $\osz$ dibyarons, and triple lines tri-nucleons.}
\end{figure}

The LO triton vertex function is the solution to a set of coupled integral equations shown in Fig.~\ref{fig:GirrLO}~\cite{Vanasse:2015fph}.
The coupled set of integral equations can be written as
\begin{align}
&\G_{t}^{(\mathrm{LO})}(p)=1+\frac{1}{\pi}\int_{0}^{\Lambda}dqq^{2}R^{(\mathrm{LO})}(q,p,E)\left\{D_{t}(q,E)\G_{t}^{(\mathrm{LO})}(q)+3D_{s}(q,E)\G_{s}^{(\mathrm{LO})}(q)\right\}\nonumber\\
&\G_{s}^{(\mathrm{LO})}(p)=1+\frac{1}{\pi}\int_{0}^{\Lambda}dqq^{2}R^{(\mathrm{LO})}(q,p,E)\left\{3D_{t}(q,E)\G_{t}^{(\mathrm{LO})}(q)+D_{s}(q,E)\G_{s}^{(\mathrm{LO})}(q)\right\},
\label{eq:vertexfunctionequationLO}
\end{align}
where
\begin{equation}
R^{(\mathrm{LO})}(q,p,E)=\frac{1}{qp}Q_{0}\left(\frac{q^{2}+p^{2}-M_{N}E}{qp}\right),
\end{equation}
and the dibaryon propagators are defined by
\begin{equation}
\label{eq:dibprop}
D_{\{t,s\}}(q,E)=\frac{1}{\sqrt{\frac{3}{4}q^{2}-M_{N}E}-\gamma_{\{t,s\}}}.
\end{equation}
Vertex functions are equivalent to Faddeev components. From them, the triton wave function can be reconstructed. For further details see Refs.~\cite{Amado:1966,Ji:2014wta,Vanasse:2015fph}. The superscripts designate that we refer here to quantities that are LO in the \eftnopi power counting.  $Q_{0}(a)$ is a Legendre function of the second kind given by 
\begin{equation}
Q_{0}(a)=\frac{1}{2}\ln\left(\frac{1+a}{a-1}\right).
\end{equation}
The binding energy, $E=-B$, where the numerical value chosen for $B$ is discussed in the next section.  In order to investigate the consequences of the Wigner-SU(4) limit in the three-body system it is convenient to rewrite the LO triton vertex function in a Wigner-SU(4) basis, which is defined by
\begin{equation}
\G_{+}^\mathrm{(LO)}(p)=\G_{t}^{(\mathrm{LO})}(p)+\G_{s}^{(\mathrm{LO})}(p)\quad,\quad\G_{-}^\mathrm{(LO)}(p)=\G_{t}^{(\mathrm{LO})}(p)-\G_{s}^{(\mathrm{LO})}(p).
\label{eq:Wignerbasis}
\end{equation}
In this basis it is necessary to take the sum and difference of the $^3$S$_1$ and $^1$S$_0$ dibaryon propagators of Eq. (\ref{eq:dibprop}).  Defining $\ya=(\gamma_{t}+\gamma_{s})/2$ and $\delta=(\gamma_{t}-\gamma_{s})/2$ the sum of dibyaron propagators is
\begin{align}
&\frac{1}{\sqrt{\frac{3}{4}q^{2}-M_{N}E}-\gamma_{t}}+\frac{1}{\sqrt{\frac{3}{4}q^{2}-M_{N}E}-\gamma_{s}}=\\\nonumber
&\hspace{5cm}\frac{2}{\sqrt{\frac{3}{4}q^{2}-M_{N}E}-\ya}\sum_{n=0}^{\infty}\frac{\delta^{2n}}{\left(\sqrt{\frac{3}{4}q^{2}-M_{N}E}-\ya\right)^{2n}}
\end{align}
and the difference
\begin{align}
&\frac{1}{\sqrt{\frac{3}{4}q^{2}-M_{N}E}-\gamma_{t}}-\frac{1}{\sqrt{\frac{3}{4}q^{2}-M_{N}E}-\gamma_{s}}=\\\nonumber
&\hspace{5cm}\frac{2}{\sqrt{\frac{3}{4}q^{2}-M_{N}E}-\gamma_{\mathrm{AVE}}}\sum_{n=0}^{\infty}\frac{\delta^{2n+1}}{\left(\sqrt{\frac{3}{4}q^{2}-M_{N}E}-\ya\right)^{2n+1}},
\end{align}
where we have expanded in powers of $\delta$ which parametrizes the distance from the Wigner-SU(4) limit.  In addition to expanding the dibaryon propagators in powers of $\delta$, the triton vertex functions are also expanded in powers of $\delta$ via
\begin{equation}
\G_{+}^\mathrm{(LO)}(p)=\sum_{n=0}^{\infty}\G_{+}^{(2n)}(p)\delta^{2n}\quad,\quad\G_{-}^\mathrm{(LO)}(p)=\sum_{n=0}^{\infty}\G_{-}^{(2n+1)}(p)\delta^{2n+1}.
\label{eq:Wignerbasisexpansion}
\end{equation}
Eqs.~(\ref{eq:Wignerbasis})--(\ref{eq:Wignerbasisexpansion}) can then be used in Eq.~(\ref{eq:vertexfunctionequationLO}), and equating terms order-by-order in $\delta$ yields 
 the set of coupled integral equations
\begin{align}
&\Gt_{+}^{(2n)}(p)=2\delta_{0n}+D(p,E)\Gt_{-}^{(2n-1)}(p)+\frac{4}{\pi}\int_{0}^{\Lambda}dq q^{2}D(q,E)R^\mathrm{(LO)}(q,p,E)\Gt_{+}^{(2n)}(q)\\\nonumber
&\Gt_{-}^{(2n+1)}(p)=D(p,E)\Gt_{+}^{(2n)}(p)-\frac{2}{\pi}\int_{0}^{\Lambda}dq q^{2}D(q,E)R^\mathrm{(LO)}(q,p,E)\Gt_{-}^{(2n+1)}(q),
\end{align}
where
\begin{equation}
D(q,E)=\frac{1}{\sqrt{\frac{3}{4}q^{2}-M_{N}E}-\ya}.
\end{equation}
The functions $\Gt^{(n)}_{\pm}(p)$ are defined by
\begin{align}
&\Gt_{+}^{(2n)}(p)=\G_{+}^{(2n)}(p)+D(p,E)\Gt_{-}^{(2n-1)}(p)\\\nonumber
&\Gt_{-}^{(2n+1)}(p)=\G_{-}^{(2n+1)}(p)+D(p,E)\Gt_{+}^{(2n)}(p).
\end{align}
Writing things in terms of $\Gt$'s, rather than $\G$'s, means that the equations simplify considerably and the correction at a given order only depends on the order immediately preceeding it, and not all orders preceeding it.  For $n=0$ we note that $\Gt_{-}^{(2n-1)}(p)=0$ and therefore $\Gt^{(0)}_{+}(p)=\G^{(0)}_{+}(p)$.  Also in the limit $\delta=0$ only the $\G^{(0)}_{+}(p)$ term gives a non-zero contribution and its integral equation is equivalent to that for three bosons~\cite{Bedaque:1999ve}.

In order to properly normalize the triton vertex function it must be multiplied by the triton wavefunction renormalization which is given by
\begin{equation}
Z_{\psi}=\frac{1}{\Sigma'(E)},
\end{equation}
where $\Sigma(E)$ is the triton self energy 
\begin{align}
\Sigma(E)=\frac{1}{2\pi}\int_{0}^{\Lambda}dqq^{2}\left\{\frac{1}{\sqrt{\frac{3}{4}q^{2}-M_{N}E}-\gamma_{t}}\G_{t}(q)+\frac{1}{\sqrt{\frac{3}{4}q^{2}-M_{N}E}-\gamma_{s}}\G_{s}(q)\right\}.
\end{align}
Again expanding the dibaryon propagators and the triton vertex functions in powers of $\delta$ we find that only even powers of $\delta$ enter in the expansion of $\Sigma$:
\begin{equation}
\Sigma(E)=\sum_{n=0}^{\infty}\Sigma^{(2n)}(E)\delta^{2n},
\end{equation}
where
\begin{equation}
\Sigma^{(2n)}(E)=\frac{1}{2\pi}\int_{0}^{\Lambda}dqq^{2}D(q)\Gt_{+}^{(2n)}(q).
\end{equation}
Thus the triton wavefunction renormalization in the $\delta$ expansion is given by 
\begin{equation}
Z_{\psi}=\frac{1}{\Sigma'(E)}=\frac{1}{{\Sigma^{(0)}}'(E)}-\frac{{\Sigma^{(2)}}'(E)}{({\Sigma^{(0)}}'(E))^{2}}+\cdots
\end{equation}

\subsection{Range corrections}
The $O(r)$ (NLO in the nuclear-force's range) correction to the triton vertex function in the $Z$-parametrization is given by~\cite{Vanasse:2015fph}
\begin{align}
&\G_{t}^{( \mathrm{NLO})}(p)=\G_{t}^{( \mathrm{LO})}(p)R_{t}^{( \mathrm{NLO})}(p,E)+\frac{1}{\pi}\int_{0}^{\Lambda}dqq^{2}D_{t}(q,E)R^{(\mathrm{LO})}(q,p,E)\G_{t}^{( \mathrm{NLO})}(q)\\[.1cm]\nonumber
&\hspace{4.3cm}+\frac{3}{\pi}\int_{0}^{\Lambda}dqq^{2}D_{s}(q,E)R^{(\mathrm{LO})}(q,p,E)\G_{s}^{( \mathrm{NLO})}(q)\\\nonumber
&\G_{s}^{( \mathrm{NLO})}(p)=\G_{s}^{( \mathrm{LO})}(p)R_{s}^{( \mathrm{NLO})}(p,E)+\frac{3}{\pi}\int_{0}^{\Lambda}dqq^{2}D_{t}(q,E)R^{(\mathrm{LO})}(q,p,E)\G_{t}^{( \mathrm{NLO})}(q)\\[.1cm]\nonumber
&\hspace{4.4cm}+\frac{1}{\pi}\int_{0}^{\Lambda}dqq^{2}D_{s}(q,E)R^{(\mathrm{LO})}(q,p,E)\G_{s}^{( \mathrm{NLO})}(q),
\end{align}
where
\begin{equation}
R_{\{t,s\}}^{( \mathrm{NLO})}(p,E)=\frac{Z_{\{t,s\}}-1}{2\gamma_{\{t,s\}}}\left(\gamma_{\{t,s\}}+\sqrt{\frac{3}{4}p^{2}-M_{N}E}\,\right).
\end{equation}
$Z_{t}=1.6908$ ($Z_{s}=.9015$) is the residue at the $^3$S$_1$ ($^1$S$_0$) channel pole~\cite{Griesshammer:2004pe,Phillips:1999hh}.  The residues $Z_{s}$ and $Z_{t}$ are equal in the Wigner-SU(4) symmetric limit.  Expanding these equations in $\delta$ gives the $\mathcal{O}(r\delta^{0})$ term\footnote{Note, when expanding in powers of $\delta$ the $\gamma_{\{t,s\}}$ in the expression $(Z_{\{t,s\}}-1)/(2\gamma_{\{t,s\}})$ is not expanded, because this whole quantity is taken as the range correction.}
\begin{align}
\G_{+}^{( \mathrm{NLO})}(p)=\G_{+}^{(0)}(p)R_{+}^{( \mathrm{NLO})}(p,E)+\frac{4}{\pi}\int_{0}^{\Lambda}dqq^{2}D(q,E)R^{(\mathrm{LO})}(q,p,E)\G_{+}^{( \mathrm{NLO})}(q),
\end{align}
where $\G_{+}^{( \mathrm{NLO})}(p)$ is the NLO-in-range-but-LO-in-Wigner correction ($\mathcal{O}(r\delta^{0})$) to $\G_{+}(p)$ and we have dropped the part of the range correction that breaks Wigner-SU(4) symmetry.  The Wigner-SU(4)-symmetric part of the range correction involves the function $R_{+}^{( \mathrm{NLO})}(p,E)$, defined as
\begin{equation}
R_{+}^{( \mathrm{NLO})}(p,E)=\rho_{\mathrm{AVE}}\left(\ya+\sqrt{\frac{3}{4}p^{2}-M_{N}E}\,\right),
\end{equation}
where
\begin{equation}
\rho_{\mathrm{AVE}}=\frac{1}{2}\left(\frac{Z_{t}-1}{2\gamma_{t}}+\frac{Z_{s}-1}{2\gamma_{s}}\right).
\end{equation}
This means that, for the $\mathcal{O}(r)$ correction, in addition to expanding in powers of $\delta$, we also expand in powers of
\begin{equation}
\delta_{r}=\frac{1}{2}\left(\frac{Z_{t}-1}{2\gamma_{t}}-\frac{Z_{s}-1}{2\gamma_{s}}\right),
\end{equation}
and the equations derived here are $\mathcal{O}(\delta_{r}^{0})$. 

\section{\label{sec:binding}Binding Energy}

\begin{figure}[hbt]
\includegraphics[width=110mm]{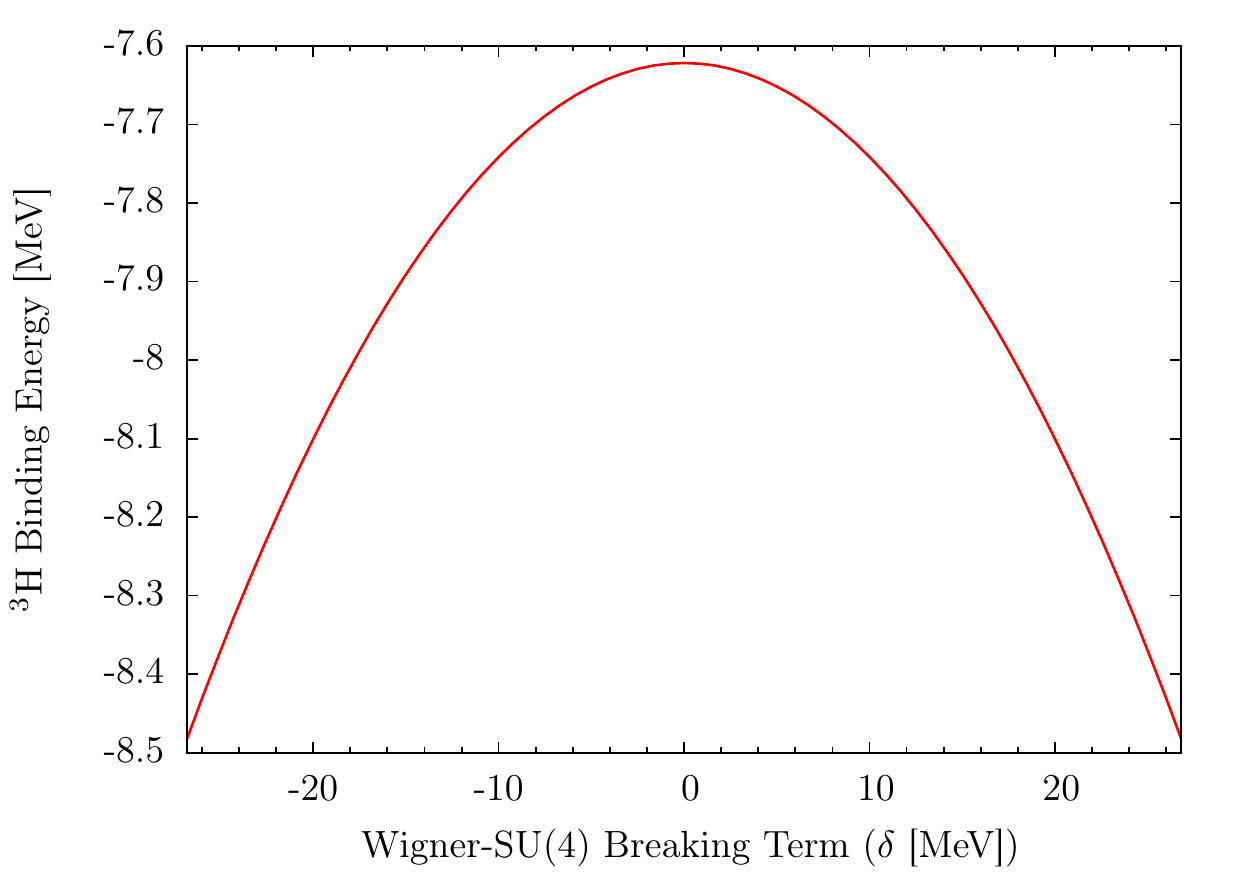}
\caption{\label{fig:E-dependence}Binding energy of the triton as a function of the Wigner-SU(4)-breaking parameter $\delta$, where the three-body force is fit to the triton binding energy at the physical value of $\delta$.  The same three-body force is used for all other values of $\delta$, and Wigner-SU(4) breaking is treated non-perturbatively.}  
\end{figure}

To understand Wigner-SU(4) breaking in the three-body system we first investigate its effects on the triton binding energy.  We do this at LO in the \eftnopi expansion. Fig.~\ref{fig:E-dependence} plots the binding energy of the triton as a function of the Wigner-SU(4) breaking parameter $\delta$, with Wigner-SU(4) breaking treated nonperturbatively.
In this calculation we employ a three-body force that is independent of $\delta$, and is fixed so as to reproduce the triton binding energy, $B_{\jjvH}=8.48$~MeV, at the physical value of $\delta=26.80$~MeV, which corresponds to the right edge of Fig.~\ref{fig:E-dependence}.  The difference between the binding energy at the physical $\delta$ and in the Wigner-SU(4) limit, $\delta=0$, is only 11\%.  The shape of the curve is essentially quadratic, demonstrating that the first Wigner-SU(4) correction to the binding energy comes in at $\mathcal{O}(\delta^{2})$.  This should come as no surprise: the vertex functions are SU(4) symmetric at leading order in the expansion in powers of $\delta$, and so the insertion of an SU(4)-breaking correction between them must yield zero. This, indeed, is why the self energy $\Sigma(E)$, has no term of $\mathcal{O}(\delta)$. 

Since in this paper we expand all observables around the Wigner-SU(4) limit, all our remaining calculations here are carried out with the binding energy chosen to have its $\delta=0$ value, $B=7.62$ MeV. This corresponds to using the same three-body force that was used to generate Fig.~\ref{fig:E-dependence}.

\section{\label{sec:radii}Charge and Matter Radii}

\subsection{Relations between radii under Wigner-SU(4) symmetry}

In the absence of Coulomb, and assuming isospin is a conserved symmetry, $\jjvHe$ is the isospin mirror of $\jjvH$.  Therefore,  the proton radius of $\jjvHe$ is the neutron radius of $\jjvH$ and vice versa.  Using this fact it is straightforward to show that for the $\jjvH$ and $\jjvHe$ wavefunctions
\begin{equation}
\label{eq:tritonradius}
\left<\!\jjvH\Bigg{|}\sum_{i}\tau_{3}^{(i)}\vect{x}_{i}^{2}\Bigg{|}\jjvH\!\right>=\left<r_{\jjvH}^{2}\right>-2\left<r_{\jjvHe}^{2}\right>,
\end{equation}
and
\begin{equation}
\left<\!\jjvHe\Bigg{|}\sum_{i}\tau_{3}^{(i)}\vect{x}_{i}^{2}\Bigg{|}\jjvHe\!\right>=2\left<r_{\jjvHe}^{2}\right>-\left<r_{\jjvH}^{2}\right>,
\end{equation}
where $\left<r_{\jjvH}^{2}\right>$ and $\left<r_{\jjvHe}^{2}\right>$ are the $\jjvH$ and $\jjvHe$ point charge radii squared respectively and ``$i$" sums over the nucleons.  In the Wigner-SU(4) limit the wavefunction is spatially symmetric such that 
\begin{equation}
\label{eq:linchpin}
\left<\!{}^{A}Z\Bigg{|}\sum_{i}\tau_{3}^{(i)}\vect{x}_{i}^{2}\Bigg{|}{}^{A}Z\!\right>=\frac{1}{3}\left<\!{}^{A}Z\Bigg{|}2T_{3}\sum_{i}\vect{x}_{i}^{2}\Bigg{|}{}^{A}Z\!\right>,
\end{equation}
where $\left.|{}^AZ\right>$ is either the $\jjvH$ or $\jjvHe$ wavefunction, and $T_{3}$ the operator for isospin in the $z$-direction on these wavefunctions.  (For a proof of this statment see Appendix~\ref{app:linchpin}.)  Noting that 
\begin{equation}
\label{eq:scalar}
\left<\!\jjvH\Bigg{|}\sum_{i}\vect{x}_{i}^{2}\Bigg{|}\jjvH\!\right>=2\left<r_{\jjvHe}^{2}\right>+\left<r_{\jjvH}^{2}\right>,
\end{equation}
and using Eqs.~(\ref{eq:linchpin}) and (\ref{eq:tritonradius}) we find

\begin{equation}
-\frac{1}{3}\left(2\left<r_{\jjvHe}^{2}\right>+\left<r_{\jjvH}^{2}\right>\right)=\left<r_{\jjvH}^{2}\right>-2\left<r_{\jjvHe}^{2}\right>.
\end{equation}
Solving this gives $\left<r_{\jjvH}^{2}\right>=\left<r_{\jjvHe}^{2}\right>$, and therefore in the Wigner-SU(4) limit the charge radii of $\jjvH$ and $\jjvHe$ are equivalent.  In addition the point matter radii for $\jjvH$ and $\jjvHe$ will be the same and equivalent to their point charge radii.  

Assuming that Wigner-SU(4) corrections are kept to all orders Eq.~(\ref{eq:scalar}) still holds.  Therefore, considering $\mathcal{O}(\delta)$ corrections Eq~(\ref{eq:scalar}) gives
\begin{equation}
\label{eq:deltaone}
\left<\!\jjvH\Bigg{|}\sum_{i}\vect{x}_{i}^{2}\Bigg{|}\delta\jjvH\!\right>+\left<\!\delta\jjvH\Bigg{|}\sum_{i}\vect{x}_{i}^{2}\Bigg{|}\jjvH\!\right>=2\left<r_{\jjvHe}^{2}\right>_{\delta}+\left<r_{\jjvH}^{2}\right>_{\delta},
\end{equation}
where $\left.|\delta\jjvH\right>$ is the first order Wigner-correction to the $\jjvH$ wavefunction and $\left<r_{\jjvH}^{2}\right>_{\delta}$ and $\left<r_{\jjvHe}^{2}\right>_{\delta}$ are the first order Wigner-corrections to the $\jjvH$ and $\jjvHe$ charge radii squared respectively.  This relationship is exactly the same for the $\jjvHe$ wavefunctions.  The quantity
\begin{equation}
\label{eq:deltatwo}
\sum_{Z}\left<{}^{A}Z\Bigg{|}\sum_{i}\vect{x}_{i}^{2}\Bigg{|}\delta{}^{A}Z\right>=0,
\end{equation}
where the sum over ``$Z$" simply sums both the $\jjvH$ and $\jjvHe$ wavefunctions.  Taking the sum over ``$Z$" makes this quantity a Wigner-SU(4) scalar, but it has one insertion of an operator that breaks Wigner-SU(4) symmetry and therefore must be zero.  Combining Eqs.~(\ref{eq:deltaone}) and (\ref{eq:deltatwo}) gives
\begin{equation}
4\left<r_{\jjvHe}^{2}\right>_{\delta}+2\left<r_{\jjvH}^{2}\right>_{\delta}=0.
\end{equation}
From this it follows that that the $\mathcal{O}(\delta)$ correction to the $\jjvH$ point charge radius squared is twice as large and has the opposite sign as the $\mathcal{O}(\delta)$ correction to the $\jjvHe$ point charge radius squared.  This relationship can also be proven using the identities in Ref.~\cite{Vanasse:2015fph} and expanding them to $\mathcal{O}(\delta)$.  However, this method is long and tedious.

\subsection{Results}

To obtain the triton charge radius in the Wigner-SU(4) limit the results of Ref.~\cite{Vanasse:2015fph} can simply be recalculated setting $\gamma_{t}=\gamma_{s}=\ya$ and $(Z_{t}-1)/2\gamma_{t}=(Z_{s}-1)/2\gamma_{s}=\rho_{\mathrm{AVE}}$.  A second approach is to take the analytical expressions in Ref.~\cite{Vanasse:2015fph} and expand them about the Wigner-limit to $\mathcal{O}(\delta)$.  This allows calculation of the $\mathcal{O}(\delta)$ correction and the calculation of the triton charge radius in the Wigner-SU(4) limit using only the triton vertex functions $\G_{+}^{(0)}(p)$ and $\G_{-}^{(1)}(p)$.  Both approaches give the same result in the limit $\delta=0$.

\begin{figure}[hbt]
\includegraphics[width=110mm]{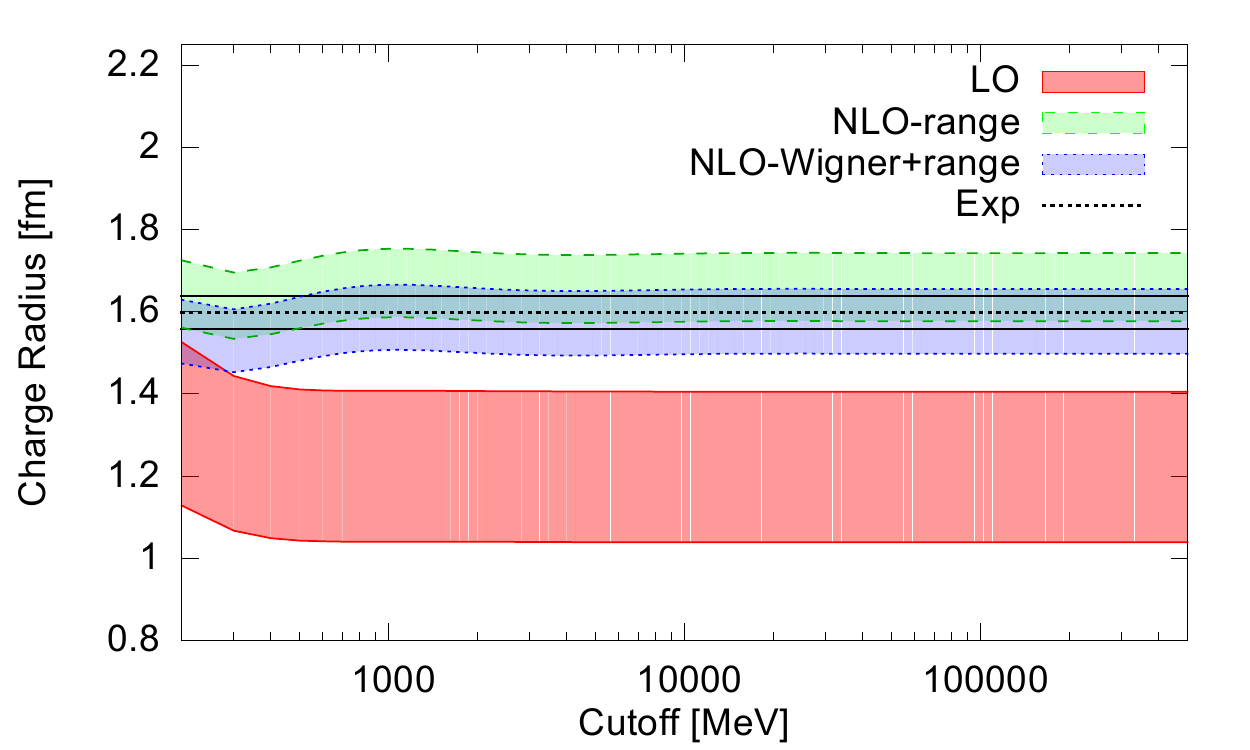}
\caption{\label{fig:ChargeRadius}Plot of cutofff dependence of LO, NLO, and $\mathcal{O}(r\!+\!\delta)$ prediction for the triton point charge radius.  The pink band corresponds to a 15\% error estimate about the LO central value, the green band to a 5\% error about the NLO central value, and the blue band a 5\% error estimate about the $\mathcal{O}(r\!+\!\delta)$ value.  The dotted black line is the experimental value for the triton point charge radius of $1.5978\pm0.040$~fm and the solid black lines about it its error~\cite{Angeli201369}.}
\end{figure}

We compute the triton point charge radius at LO ($\mathcal{O}(r^{0}\delta^{0})$), NLO ($\mathcal{O}(r\delta^{0})$), and $\mathcal{O}(r\!+\!\delta)$, where the last calculation involves the addition of both a single range insertion and a single Wigner-SU(4)-breaking insertion, but only considered separately, not in combination.  Cutoff dependence of these three different results is displayed in Fig.~\ref{fig:ChargeRadius}.
All orders of the triton point charge radius considered here converge as a function of cutoff, and are therefore properly renormalized.  The LO triton point charge radius is $1.22$~fm, the NLO value $1.66$~fm, and the $\mathcal{O}(r\!+\!\delta)$ value $1.58$~fm. The experimental value for the triton point charge radius is $1.5978\pm0.040$~fm~\cite{Angeli201369}, which agrees well with our $\mathcal{O}(r\!+\!\delta)$ calculation. When Wigner-SU(4) breaking is included to all orders, i.e., the physical values of $\gamma_s$ and $\gamma_t$, and the physical triton binding energy, $B_{\jjvH}=8.48$ MeV, are employed, at LO (NLO) in \eftnopi the triton point charge radius is
 $1.14$~fm  ($1.59$~fm)~\cite{Vanasse:2015fph}.    $\jjvHe$ has an experimental point charge radius of $1.77527\pm0.0054$~fm~\cite{Angeli201369}. This is about 7\% away from the NLO-in-range-but-SU(4)-symmetric prediction of $1.66$~fm.  As already noted, the Wigner-SU(4)-breaking correction for the $\jjvHe$ point charge radius squared is half that for the $\jjvH$ point charge radius squared and of opposite sign.  Therefore, the $\mathcal{O}(r\!+\!\delta)$ $\jjvHe$ point charge radius is $1.70$~fm, about 4\% away from the experimental value.  

The error due to missing range corrections is about 10\%. The dominant, SU(4)-symmetric, part of this correction  will affect the $\jjvH$ and $\jjvHe$ charge radii equally.  In contrast, the effects of the Coulomb interaction, not included here, will affect only the charge radius of $\jjvHe$. We estimate this effect to be of order $\alpha M_{N}/\kappa_t$ (where $\kappa_t=\sqrt{M_{N}B_{\jjvH}}$ is the binding momentum of the triton), which is about 8\%.  Meanwhile, the uncertainty due to Wigner-SU(4) breaking in the \NN effective ranges is naively 3\% since
\begin{equation}
\frac{\delta_r}{a} \approx \left\{\left(\frac{Z_{t}-1}{2\gamma_{t}}-\frac{Z_{s}-1}{2\gamma_{s}}\right)\!\!\Big{/}\!\!\left(\frac{Z_{t}-1}{2\gamma_{t}}+\frac{Z_{s}-1}{2\gamma_{s}}\right)\right\}\left(\frac{\gamma_{t}}{m_{\pi}}\right)\sim.033.
\end{equation}
Terms of $\mathcal{O}(r \delta)$ are also omitted. These could also be as large as a few per cent of the individual radii, since range corrections to those are large.  Corrections that are Wigner-SU(4) odd (e.g. $\mathcal{O}(\delta)$, $\mathcal{O}(r \delta)$, and $\mathcal{O}(r \delta_{r})$) will affect only the isovector combination of trinucleon charge radii,
\begin{equation}
\left<r^{2}_{v}\right>=\frac{1}{2}\left(2\left<r^{2}_{\jjvHe}\right>-\left<r^{2}_{\jjvH}\right>\right),
\end{equation}
and give zero contribution to to the isoscalar combination:
\begin{equation}
\left<r^{2}_{s}\right>=\frac{1}{2}\left(2\left<r^{2}_{\jjvHe}\right>+\left<r^{2}_{\jjvH}\right>\right).
\end{equation}
Finally, considering the  convergence of the expansion in powers of $\delta$, e. g., the ratio between $\G_{+}^{(0)}(p)$ and $\G_{+}^{(2)}(p)$, suggests that $\mathcal{O}(\delta^2)$ effects could have perhaps a 5\% effect on the radii.

\section{\label{sec:expansion}Convergence of the Wigner-SU(4) Expansion}

In order to assess the efficacy of expanding about the Wigner-SU(4) limit we plot the relative error of the triton vertex function with the breaking of Wigner-SU(4) symmetry in the \NN scattering lengths included to all orders, as compared with that obtained when this source of Wigner-SU(4) symmetry breaking is treated perturbatively order-by-order in $\delta$.  

\begin{figure}[hbt!]
\includegraphics[width=110mm]{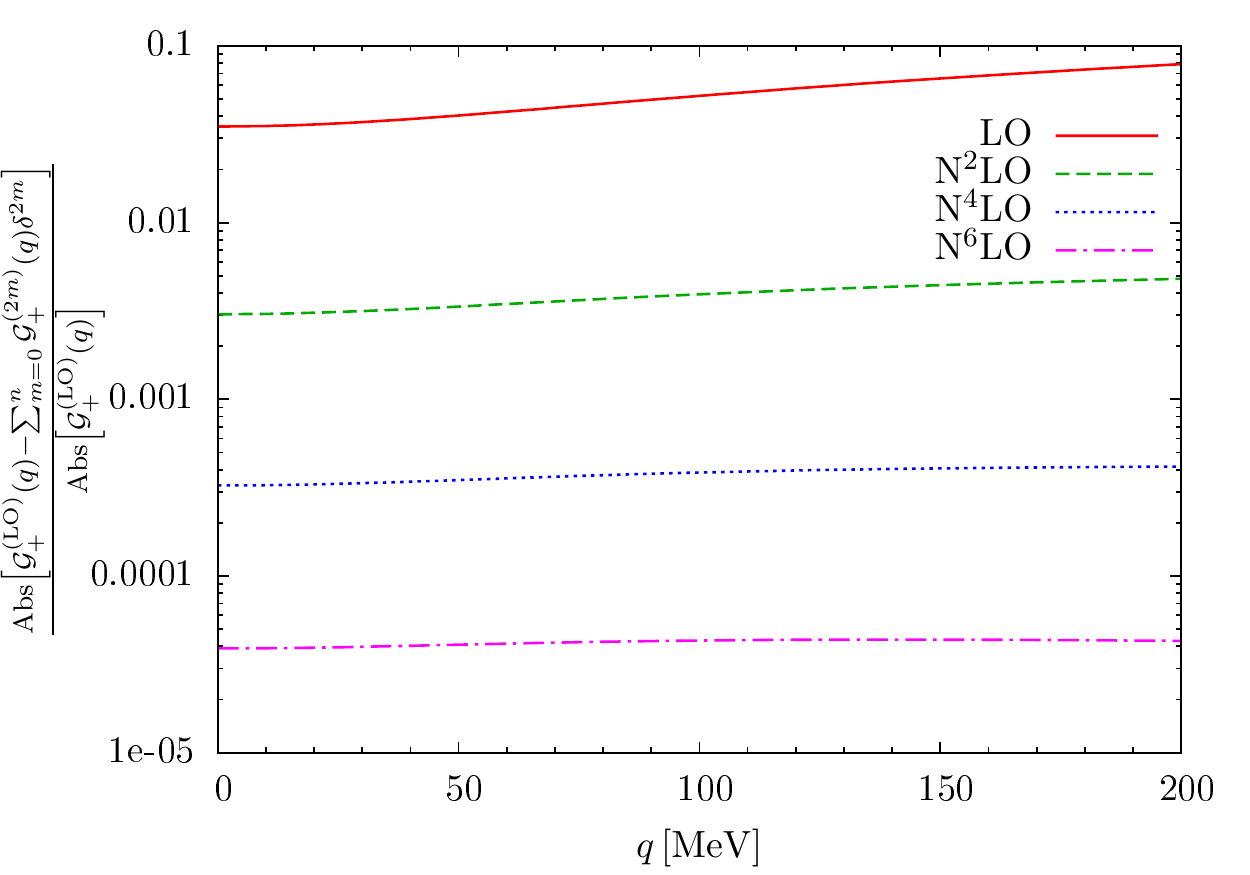}
\caption{\label{fig:G-plus}Plot of relative difference between $\G_{+}(q)$ and $\sum_{m=0}^n \G_{+}^{(2m)}(q)\delta^{2m}$ to $\mathcal{O}(\delta^{6})$.  The relative error is plotted over the range $q=0-200$~MeV and the data is for the cutoff $\Lambda=51286$~MeV.  The LO ($\mathcal{O}(\delta^{0})$) result is given by the solid red curve, the N$^{2}$LO ($\mathcal{O}(\delta^{2})$) result by the long-dashed green curve, the N$^{4}$LO ($\mathcal{O}(\delta^{4})$) result by the short-dashed blue curve, and the N$^{6}$LO ($\mathcal{O}(\delta^{6})$) result by the short-long-dashed purple curve.}
\end{figure}

\begin{figure}[hbt!]
\includegraphics[width=110mm]{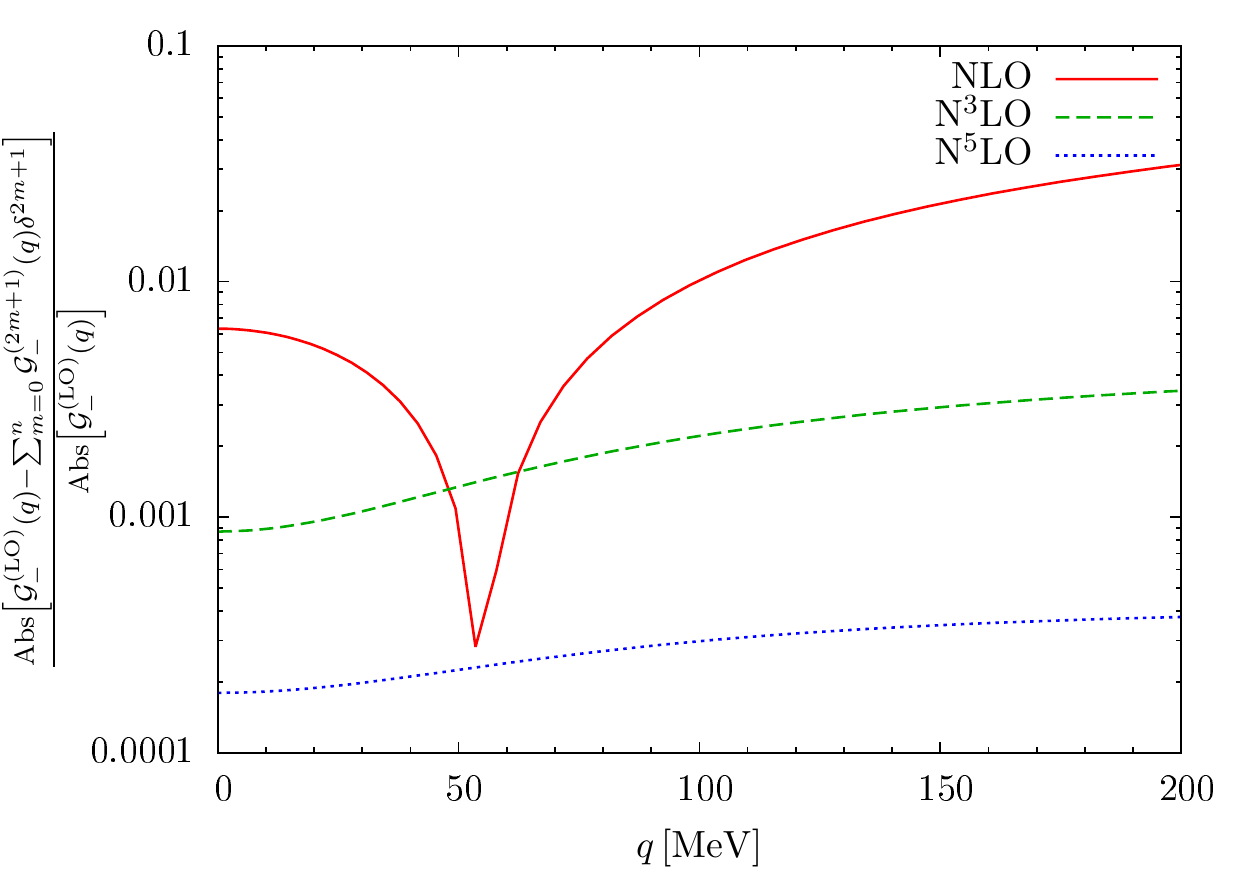}
\caption{\label{fig:G-minus}Plot of relative {error of}{difference between} $\G_{-}^{(\mathrm{LO})}(q)$ and $\sum_{m=0}^n \G_{-}^{(2m+1)}(q)\delta^{2m+1}$ to $\mathcal{O}(\delta^{5})$.  The relative error is plotted over the range $q=0-200$~MeV and the data is for the cutoff $\Lambda=51286$~MeV. The NLO ($\mathcal{O}(\delta)$) result is given by the solid red curve, the N$^{3}$LO ($\mathcal{O}(\delta^{3})$) result by the long-dashed green curve, and the N$^{5}$LO ($\mathcal{O}(\delta^{5})$) result given by the short-dashed blue curve.}
\end{figure}

Figure~\ref{fig:G-plus} shows the relative error of the cumulative sum in the expansion in powers of $\delta$, $\sum_{m=0}^n \G_{+}^{(2m)}(q)\delta^{2m}$, compared to $\G_+^{(\mathrm{LO})}(q)$, up to $\mathcal{O}(\delta^{6})$, over a range of momenta that essentially corresponds to the domain of validity of \eftnopi, $q=0-200$~MeV. The data is chosen at the cutoff $\Lambda=51286$~MeV; by this cutoff all results have effectively converged as a function of $\Lambda$.  Order-by-order convergence in the $\delta$ expansion can clearly be seen in the relative error. 

\begin{figure}[hbt]
\includegraphics[width=110mm]{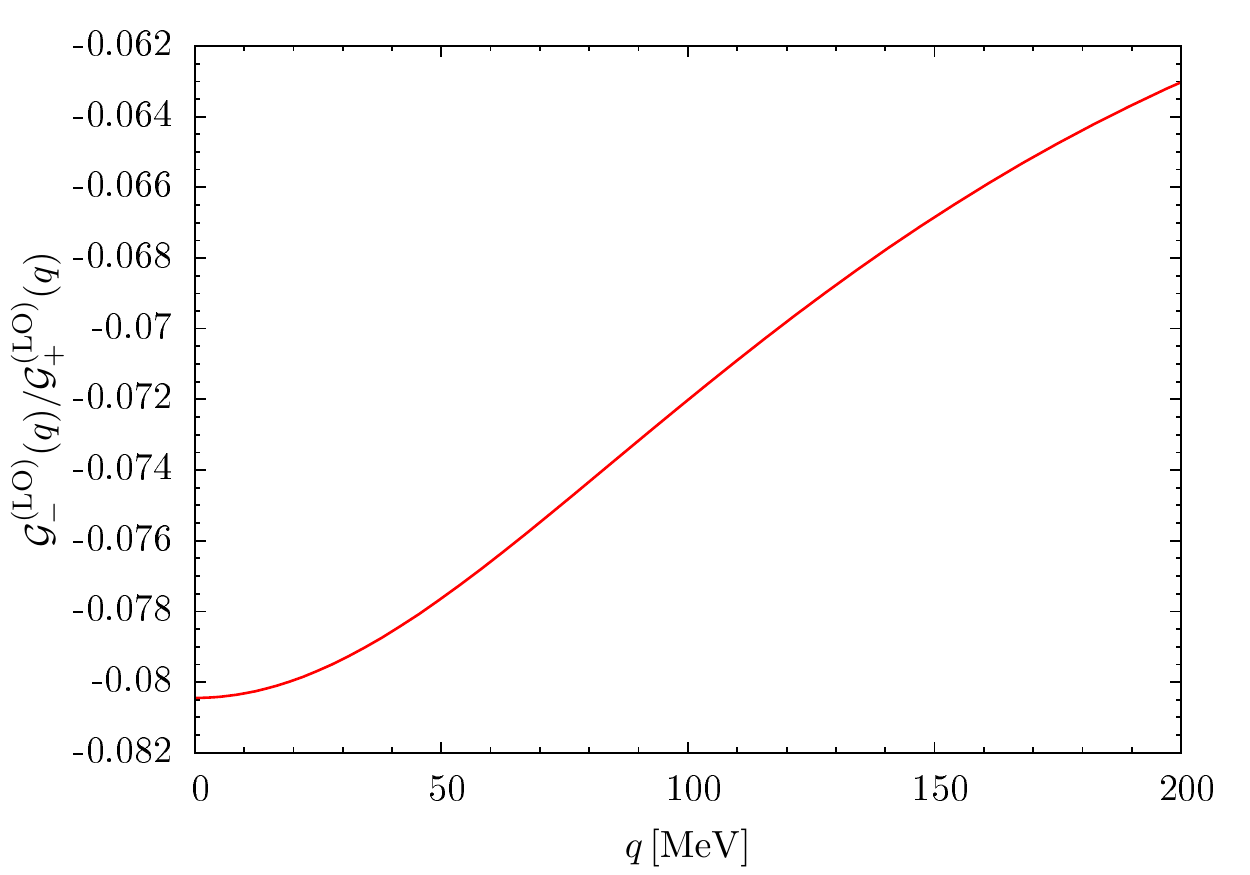}
\caption{\label{fig:GminusdivGplus} The ratio of $\G_{-}^{(\mathrm{LO})}(q)$ and $\G_{+}^{(\mathrm{LO})}(q)$, again computed for a cutoff of $\Lambda=51286$ MeV.}
\end{figure}

In Fig.~\ref{fig:G-minus} the relative difference between $\G_{-}^{(\mathrm{LO})}(q)$  and the cumulative sum $\sum_{m=0}^n \G_{-}^{(2m+1)}(q)\delta^{2m+1}$ is shown for the same range of momenta and the same cutoff $\Lambda$, up to $\mathcal{O}(\delta^{5})$.  Again, order-by-order convergence is clearly observed.  The dip at NLO merely corresponds to the fact that $\G_{-}^{(\mathrm{LO})}(q)$ and $\G_{-}^{(1)}(q)\delta$ cross each other at a momentum $\approx \gamma_t$, i.e., about 50 MeV.  Finally, we compare the size of $\G_{-}^{(\mathrm{LO})}(q)$ and $\G_{+}^{(\mathrm{LO})}(q)$, see Fig.~\ref{fig:GminusdivGplus}. $\G_{-}^{(\mathrm{LO})}(q)$ is at most 8\% of $\G_{+}^{(\mathrm{LO})}(q)$ over the entire momentum region of interest.

These results support the claim that an expansion about the Wigner-SU(4) limit converges rapidly. We recognize, of course, that triton vertex functions are not observables. However, since the construction of any three-nucleon bound-state property in \eftnopi will involve these non-perturbative objects, the fact that they converge rapidly in the $\delta$ expansion suggests that the expansion will generally be successful for three-nucleon bound-state observables.

\section{\label{sec:unitary}Comments on the unitary limit}

K\"onig {\it et al.}  have recently argued that the binding energies of nuclei up to $A=4$ can be understood in an expansion about the unitary limit, where $\gamma_s=\gamma_t=0$~\cite{Konig:2016utl}. The unitary limit is clearly a special case of Wigner-SU(4) symmetry; taking $\gamma_s=\gamma_t=0$ enlarges the symmetry group still further, since the discrete scale invariance of \eftnopi at LO then relates all the unitary-limit Efimov states by a fixed rescaling.  In the case of finite scattering lengths the Efimov spectrum still possesses discrete scale invariance, but a particular Efimov state is related to others at a different \NN scattering length~\cite{Braaten:2004rn}.  As we have done here, K\"onig {\it et al.} fix the size of the three-body force to reproduce the binding energy of the physical triton. They demonstrate that the binding-energy difference of ${}^3$He and ${}^3$H remains well predicted in the unitary limit (cf. Refs.~\cite{Kirscher:2011zn,Konig:2015aka}). They also show that the alpha particle, while overbound by about 10 MeV at exact unitarity, attains almost exactly its experimental binding energy (28.30 MeV) once first-order corrections in the expansion in $\gamma_{t}$ are included. 

It is straightforward for us to take the limit $\ya \rightarrow 0$ in our results and so obtain point charge radii for three-nucleon bound states in the unitary limit. At leading order in the range expansion this gives $\langle r^2 \rangle_{\rm pt}^{1/2}=1.10$ fm (for $B=7.62$ MeV), in accord with the analytic result~\cite{Braaten:2004rn}\footnote{In fact, Ref.~\cite{Braaten:2004rn} quotes this as the result for the matter radius in the unitary limit for three equal-mass particles. However, in that limit the symmetry of the three-body wave function leads to equal charge and matter radii.}
\begin{equation}
\langle r^2 \rangle_{\rm pt} M_N B=\frac{(1+s_{0})^{2}}{9}\approx.224.
\label{eq:unitarylimitradius}
\end{equation}
This lends support to the argument of K\"onig {\it et al.}, since it is within 10\% of either the Wigner-SU(4) limit result quoted above, or the full LO \eftnopi answer of 1.14 fm~\cite{Vanasse:2015fph}.

Adding corrections of first order in the range of the \NN interaction corrects the radius obtained from Eq.~(\ref{eq:unitarylimitradius}) by an amount $\sim r \kappa_t$---such effects are present even though $r/a=0$ at unitarity. (Note, however, that the factors $Z_s$ and $Z_t$ remain at their LO values of one as long as we consider $\gamma_s=\gamma_t=0$~\footnote{In the unitary limit and Wigner-SU(4) limit $(Z_{\{t,s\}}-1)/(2\gamma_{\{t,s\}})\to \frac{1}{2}\rho$, where $\rho=\frac{1}{2}(\rho_{t}+\rho_{s})$, with $\rho_{t}=1.765$~fm ($\rho_{s}=2.730$~fm) being the effective range about the $^3$S$_1$ ($^1$S$_0$) pole.}.)  This shifts $\langle r^2 \rangle_{\rm pt}^{1/2}$ to 1.42 fm, i.e., the size of the range correction at unitarity is about 70\% of that when $\ya$ takes its physical value. 

\begin{table}
\begin{tabular}{|l|c|c|c|c|}
\hline
& Unitary limit & Wigner-SU(4) limit & $\mathcal{O}(\delta)$ & Full Wigner-SU(4) breaking\\
\hline \hline
LO \eftnopi ($r=0$)  & 1.10 & 1.22 & 1.08/1.19 & 1.14/1.26\\
\hline
$\mathcal{O}(r)$ & 1.42 & 1.66 & 1.58/1.70 & 1.59/1.72 \\
\hline
\end{tabular}
\caption{Anatomy of the point charge radii $\langle r^2 \rangle_{\rm pt}^{1/2}$, of three-nucleon bound states. When only one number is quoted the radii are equal for $\jjvH$ and $\jjvHe$, if two numbers are given the first is for the triton and the second for $\jjvHe$.
All numbers are in fm. Note that the lower line, third entry in the table is the $\mathcal{O}(r + \delta)$ calculation of this paper.  The ``Full Wigner-SU(4) breaking" numbers  treat Wigner-breaking in the scattering lengths and (in the second line) effective ranges nonperturbatively and use the physical triton binding energy~\cite{Vanasse:2015fph,Vanasse:2016}. The experimental evaluation of Ref.~\cite{Angeli201369} quotes 1.598(40)/1.7753(54) .}
\label{table-anatomy}
\end{table}

Table~\ref{table-anatomy} summarizes the effect that different limits in the \NN system have on the radii of the three-nucleon bound states. 
In the case of the triton we can compare to Ref.~\cite{Vanasse:2015fph}, which obtained 1.14 fm at LO in \eftnopi, but with Wigner-SU(4) breaking included to all orders in $\delta$, and 1.59 fm in a calculation that was first order in the ranges (including SU(4) breaking therein), and again had the physical values of $\gamma_s$ and $\gamma_t$. The proximity of our $\mathcal{O}(r+\delta)$ results to these is very striking.

The unitary limit seems a worse starting point---at least for radii---especially since the shift that results from range corrections is significantly underpredicted there. It may be that radii are more challenging for the expansion proposed in Ref.~\cite{Konig:2016utl}, since they are quite sensitive to infra-red physics, and the long-distance properties of the three-nucleon system in the unitary limit differ dramatically from reality: at both LO and NLO in the expansion of Ref.~\cite{Konig:2016utl} infinite towers of bound Efimov excited states occur.

\section{\label{sec:conclusions}Conclusions}

Working to $\mathcal{O}(r+\delta)$ in the range and $\delta$ expansion of \eftnopi we obtain a point charge radius for $\jjvH$ of $1.58$~fm, which agrees with the experimental number, $1.5978\pm0.040$~fm~\cite{Angeli201369}, within the experimental errors.  It also agrees with the NLO result of $1.59$~fm obtained using the physical values of the \NN scattering lengths and triton binding energy~\cite{Vanasse:2015fph}, within theoretical errors. It follows that all higher-order corrections in $\delta$ and $\delta_{r}$ must conspire to give a total correction of only $.01$~fm to $\langle r_{\jjvH}^{2} \rangle_{\rm pt}^{1/2}$. Naively $\mathcal{O}(\delta^{2})$ corrections could give 5\% of the LO Wigner SU(4)-symmetric charge radius $1.22$~fm, i.e., they should be $\approx .06$~fm.  However, at $\mathcal{O}(\delta^{2})$ there will be effects both from expanding the expressions of Ref.~\cite{Vanasse:2015fph} out to $\mathcal{O}(\delta^2)$, and from the $\mathcal{O}(\delta^2)$ shift of the three-nucleon bound state energy from $B=7.62$~MeV to the physical triton binding energy.  The small overall result of a $.01$~fm shift is probably due to a cancellation between these two classes of $\mathcal{O}(\delta^2)$ corrections.

Working to first order in both Wigner-SU(4) breaking and the \NN effective range produces a $\jjvHe$ point charge radius of $1.70$~fm, about 4\% below the experimental value of $1.77527\pm 0.0054$~fm~\cite{Angeli201369}.  The difference is mostly from missing Coulomb and higher-order range corrections, since the $\jjvHe$ charge radius with Wigner SU(4)-breaking included to all orders (including breaking in ranges), but no Coulomb effects, is $1.72$~fm at NLO in the range expansion~\cite{Vanasse:2016}.  

In this (isospin symmetric) limit $\jjvH$ and $\jjvHe$ have a common binding energy. It is thus an SU(4) scalar, and so receives no correction at $\mathcal{O}(\delta)$.  We find that $\mathcal{O}(\delta^2)$ effects make the triton 11\% less bound in the Wigner-SU(4) limit than it is at the physical value, $\delta = 27$~MeV.

$\delta$ is in fact larger than $\ya=19$ MeV, and so the rapid convergence of the expansion in powers of $\delta$ at first glance is somewhat mysterious. However, the expansion is really an expansion in powers of $\delta D(q,E)$, with $D(q,E)$ the
\eftnopi propagator for the \NN system that appears in the three-body equations. This renders the expansion around the SU(4) limit one in $(\gamma_t-\gamma_s)/\kappa_t$, with $\kappa_t =89$ MeV the binding momentum of the triton.

Examining both the three-nucleon binding energy and the relative size of the SU(4)-symmetric and SU(4)-anti-symmetric pieces of the three-nucleon vertex function, $\G_{-}^{(\mathrm{LO})}(p)/\G_{+}^{(\mathrm{LO})}(p)$, suggests that the error induced in observables through going to the Wigner-SU(4) limit will be at most 10\%.  This implies that an efficient way to account for Wigner-SU(4) breaking is to equate $\delta \sim r^2$, i.e. only compute one correction in Wigner-SU(4) breaking for every two orders in the range expansion.  Unfortunately, an $\mathcal{O}(r^2)$ calculation requires a new three-body force that must be renormalized to a three-body datum~\cite{Bedaque:2002yg,Ji:2012nj}.  Since the Wigner-SU(4) limit is not expected to work nearly as well for scattering observables that additional three-body force should be fit to a three-body bound state observable. We postpone this to future work. 

Finally, we note that the Wigner-SU(4) symmetry which emerges in \eftnopi is not obviously related to the contracted SU(4) of QCD in the limit of a large number of colors (\Nc)~\cite{Dashen:1993jt}. In the large-\Nc limit Wigner-SU(4) symmetry of nuclear forces naturally emerges~\cite{Kaplan:1995yg,Kaplan:1996rk,Phillips:2013rsa}, but this happens only at a renormalization scale $\sim \Lambda_{\mathrm{QCD}}$, whereas the SU(4) in \eftnopi emerges already for renormalization scales $\sim m_\pi$.  

\appendix 

\section{Proof of Equation~(\ref{eq:linchpin})}
\label{app:linchpin}

In order to prove Eq.~(\ref{eq:linchpin}) the spatial permutation operator is defined as $P_{ij}$.  This operator permutes the $i$th and $j$th particles in the spatial part of the wavefunction while leaving the spin and isospin parts of the wavefunction untouched.  Noting $P_{ij}^2=1$ gives
\begin{align}
&\left<\!{}^{A}Z\Bigg{|}\sum_{i}\tau_{3}^{(i)}\vect{x}_{i}^{2}\Bigg{|}{}^{A}Z\!\right> \\\nonumber
&\hspace{2cm}=\frac{1}{3}\left<\!{}^{A}Z\Bigg{|}\sum_{i}\tau_{3}^{(i)}(\vect{x}_{i}^{2}+P_{ij}^2 \vect{x}_{i}^{2}P_{ij}^2 +P_{ik}^2 \vect{x}_{i}^{2} P_{ik}^2) \Bigg{|}{}^{A}Z\!\right>,
\end{align}
where $i\neq j$, $i\neq k$, and $j\neq k$.  In the Wigner SU(4)-limit the spatial part of the tri-nucleon wavefunction is spatially symmetric since it is equivalent to that of three bosons, and is thus invariant under any spatial permutation.  Now, since the spatial permutation operator does not act on isospin it can be commuted with $\tau_{3}^{(i)}$, leading to
\begin{align}
&\left<\!{}^{A}Z\Bigg{|}\sum_{i}\tau_{3}^{(i)}\vect{x}_{i}^{2}\Bigg{|}{}^{A}Z\!\right> \\\nonumber
&\hspace{2cm}=\frac{1}{3}\left<\!{}^{A}Z\Bigg{|}\sum_{i}\tau_{3}^{(i)}(\vect{x}_{i}^{2}+P_{ij}\vect{x}_{i}^{2}P_{ij}^{\dagger}+P_{ik}\vect{x}_{i}^{2}P_{ik}^{\dagger})\Bigg{|}{}^{A}Z\!\right>,
\end{align}
which reduces finally to
\begin{align}
&\left<\!{}^{A}Z\Bigg{|}\sum_{i}\tau_{3}^{(i)}\vect{x}_{i}^{2}\Bigg{|}{}^{A}Z\!\right> \\\nonumber
&\hspace{2cm}=\frac{1}{3}\left<\!{}^{A}Z\Bigg{|}\sum_{i}\tau_{3}^{(i)}(\vect{x}_{i}^{2}+\vect{x}_{j}^{2}+\vect{x}_{k}^{2})\Bigg{|}{}^{A}Z\!\right>\\\nonumber
&\hspace{2cm}=\frac{1}{3}\left<\!{}^{A}Z\Bigg{|}2T_{3}\sum_{i}\vect{x}_{i}^{2}\Bigg{|}{}^{A}Z\!\right>.
\end{align}

\acknowledgments{We thank Shung-Ichi Ando for comments that helped us clarify the manuscript.  We are grateful to the ExtreMe Matter Institute EMMI at the GSI Helmholtz Centre for Heavy Ion Research for support as part of the Rapid Reaction Task Force, ``The systematic treatment of the Coulomb interaction in few-body systems".  We acknowledge financial support by the U.S. Department of Energy, Office of Science, Office of Nuclear Physics, under Award Number DE-FG02-93ER40756.}

\bibliographystyle{h-physrev5}

\end{document}